\documentclass[11pt,twocolumn,twoside]{article}
\usepackage{fully3d}
\graphicspath{{images/}}
\usepackage{subfig}
\usepackage{hyperref}
\usepackage{xcolor}
\hypersetup{
  colorlinks   = true, 
  urlcolor     = red, 
  linkcolor    = blue, 
  citecolor   = blue 
}

\begin{document}

\title{MBIR Training for a 2.5D DL network in X-ray CT}

\author[1]{Obaidullah~Rahman}
\author[2]{Madhuri~Nagare}
\author[1]{Ken~D.~Sauer}
\author[2]{Charles~A.~Bouman}
\author[3]{Roman~Melnyk}
\author[3]{Brian~Nett}
\author[3]{Jie~Tang}

\affil[1]{Department of Electrical Engineering,
          University of Notre Dame, USA}

\affil[2]{School of Electrical Electrical and Computer Engineering,
          Purdue University, West Lafayette, USA}

\affil[3]{General Electric Healthcare,
          Waukesha, USA}          

\maketitle
\thispagestyle{fancy}


\begin{customabstract}
In computed tomographic imaging, model based iterative reconstruction methods
have generally shown better image quality than the more traditional, faster filtered
backprojection technique. 
The cost we have to pay is that MBIR is computationally expensive. 
In this work we train a 2.5D deep learning (DL) network to mimic MBIR quality image. 
The network is realized by a modified Unet, and trained using clinical FBP and MBIR image pairs. 
We achieve the quality of MBIR images
faster and with a much smaller computation cost. 
Visually and in terms of noise power spectrum (NPS), DL-MBIR images have texture
similar to that of MBIR, with reduced noise power.
Image profile plots, NPS plots, standard deviation, etc. suggest
that the DL-MBIR images result from a successful emulation of an MBIR operator.
\end{customabstract}


\section{Introduction}

%

X-ray computed tomography has become a important tool in applications such as
healthcare diagnostics, security inspection, and non-destructive testing. 
The industry
preferred method of reconstruction is filtered backprojection (FBP) and its popularity is owed
to its speed and low computational cost. 
Iterative methods such as model-based iterative reconstruction (MBIR) generally
have better image quality than FBP and do better in limiting image artifacts \cite{thibault2007three, yu2010fast}.

MBIR is a computationally expensive and potentially slow reconstruction method since it entails repeated
forward projection of the estimated image and back projection of the sinogram residual error. 
Even with fast GPUs becoming the norm, MBIR may
take minutes compared to an FBP reconstruction that can be performed in seconds. 
The computational cost and reconstruction time have been deterents in wide
adoption of MBIR. 

In recent years, deep learning has made serious inroads in CT applications. 
It is applied in sinogram and image domains and sometimes in both. 
It has been applied in low signal correction \cite{geng2018unsupervised}, 
image denoising \cite{yang2018low,matsuura2020feature}, and metal artifact reduction \cite{ghani2019fast}.

Ziabari, et al \cite{ziabari20182} showed that a 2.5D deep neural network, with proper training, can effectively
learn a mapping from an FBP image to MBIR.
In this paper we expand on their work and study the characteristics of output from networks trained to 
simulate MBIR with a highly efficient neural network implementation.

\section{Methods}
We will first train a deep neural network, which we will, similarly to \cite{ziabari20182}, entitle DL-MBIR.
Our aim is to train the network to closely approximate MBIR images from FBP images.
The training input is FBP images and the target is MBIR images from the same data.
Let $X_{FBP}$ be the input to the network, $X_{MBIR}$ be the target, and $\sigma$ represents a hypothetical mapping
such that $\sigma: X_{MBIR} \rightarrow X_{FBP}$. 
Let $f_{DL \text{-} MBIR_Z}$ be the DL neural network with $Z$ number of input channels. During the training phase:
\begin{align}
&\hat{f}_{DL \text{-} MBIR_Z} = \operatorname*{argmin}_{f_{DL \text{-} MBIR_Z} } {\big |}{\big |} f_{DL \text{-} MBIR_Z} (X_{FBP}) - X_{MBIR} {\big |}{\big |}_2
\end{align}
$f_{DL \text{-} MBIR_Z} $ can be thought of as the inverse of $\sigma$, i.e. $f_{DL \text{-} MBIR_Z}  = \sigma^{-1}$. 
During the training phase,
the weights of $f_{DL \text{-} MBIR_Z} $ are randomly initialized and then adjusted in several iterations using error backpropagation.
Once the number of iterations is exhausted or the convergence criteria is met, the training stops.

\begin{figure}
  \centering
  \includegraphics[width=0.50\textwidth]{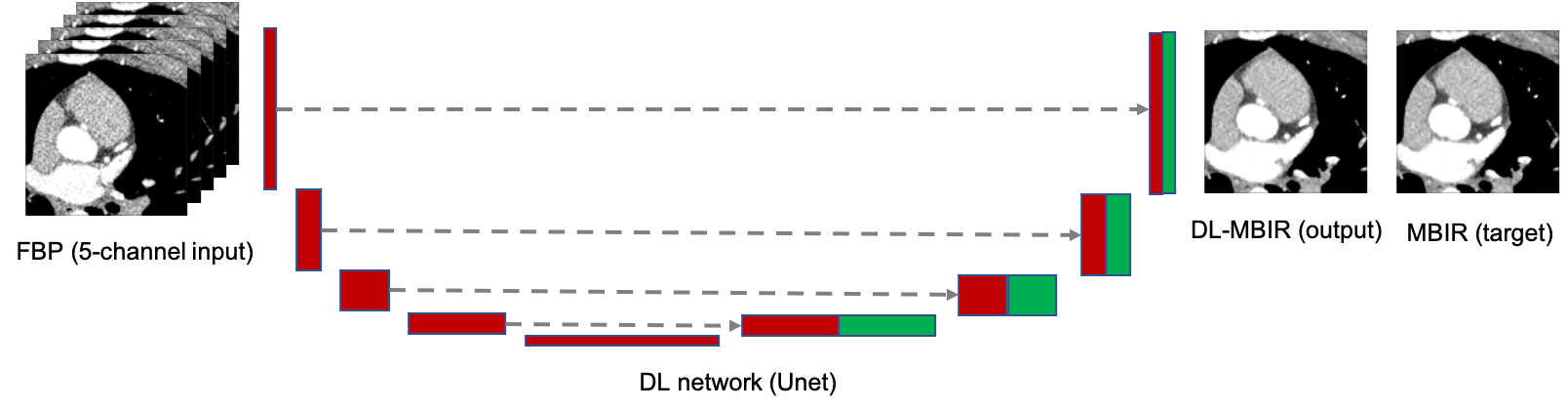}
  \caption{DL architecture and training setup. This is a modified form of Unet. Layers on the left denote the contracting path where
  features are compressed from image towards latent space but the number of features increases. Layers on the right denote the
  expanding path where feature are decompressed from latent space towards corrected image but the number of features decreases.}
  \label{fig:Unet}
\end{figure}

For training, 4 pairs of clinical exams were selected. Each pair had one FBP image volume and the corresponding MBIR volume.
Each image volume had about 200 slices, resulting in about 800 training image pairs. 
A modified version of Unet \cite{ronneberger2015u} was chosen as the network architecture. 
The learning rate was set to $0.0001$ and 2 GPUs were used. Training and inferencing were done on Tensorflow/Keras.
Training was run for 300 epochs. 3 versions of DL-MBIR were trained: $DL \text{-} MBIR_1$ was trained with inputs with
1 channel i.e. 1 axial slice, $DL \text{-} MBIR_3$ was trained with 3-channel inputs and $DL \text{-} MBIR_5$ was trained with 5-channel inputs.
Having adjacent slices in the input provides additional information to the DL network \cite{ziabari20182} and helps train it better.
Figure \ref{fig:Unet} shows the DL architecture and the training setup.

\section{Results}
\begin{figure}[h!]
  \begin{center}
	\subfloat[]{\includegraphics[width=0.45\textwidth]{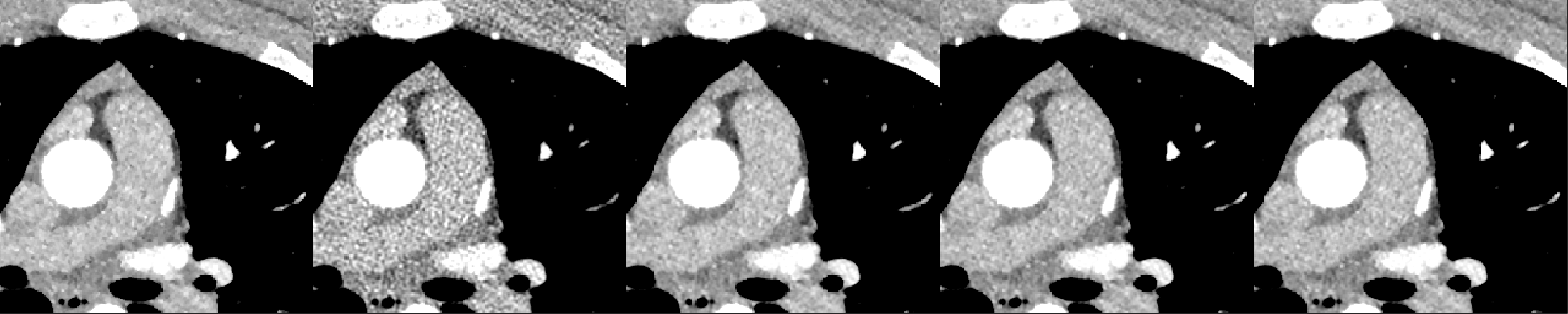}\label{fig:Image_10}}
	\hspace{0.01 cm}
	\subfloat[]{\includegraphics[width=0.45\textwidth]{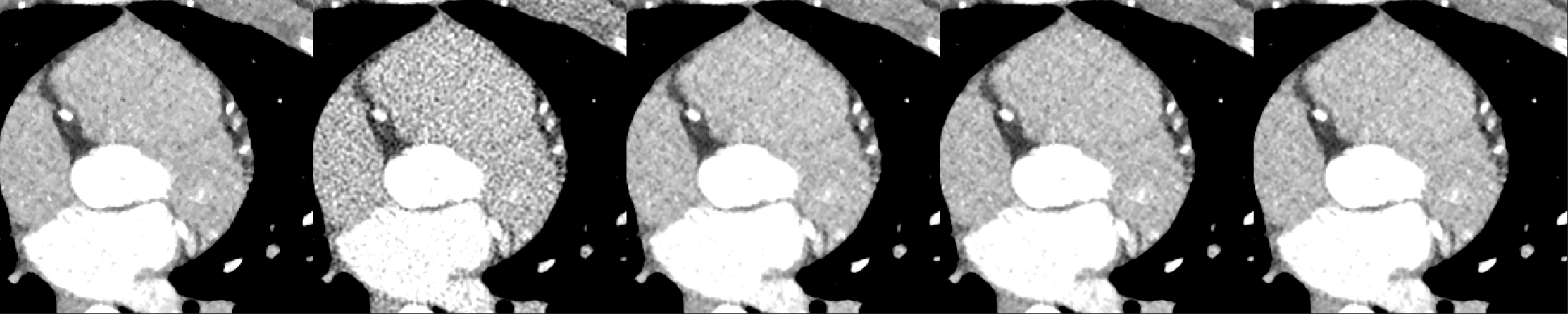}\label{fig:Image_50}}
	\hspace{0.01 cm}
	\subfloat[]{\includegraphics[width=0.45\textwidth]{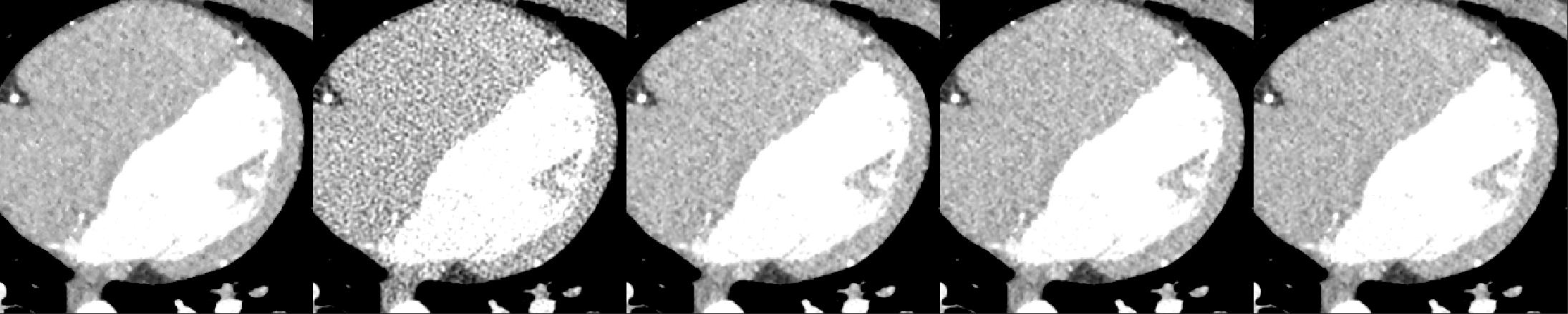}\label{fig:Image_90}}
	\hspace{0.01 cm}
	\subfloat[]{\includegraphics[width=0.45\textwidth]{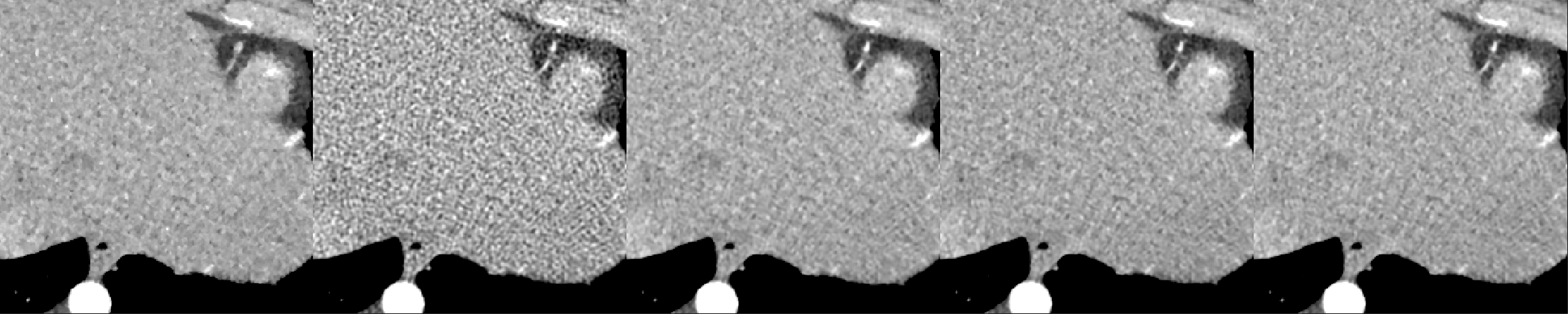}\label{fig:Image_170}}
	\caption{Reconstructed image. (left to right): $MBIR$, $FBP$, $DL \text{-} MBIR_1$, $DL \text{-} MBIR_3$, $DL \text{-} MBIR_5$.
	\label{fig:recon images}
	\protect\subref{fig:Image_10},  \protect\subref{fig:Image_50}, \protect\subref{fig:Image_90} and \protect\subref{fig:Image_170} represent different slices in the image volume. WW/WL 450/0 HU.}
  \end{center}
\end{figure}

\begin{figure}[h!]
  \begin{center}
	\subfloat[]{\includegraphics[width=0.45\textwidth]{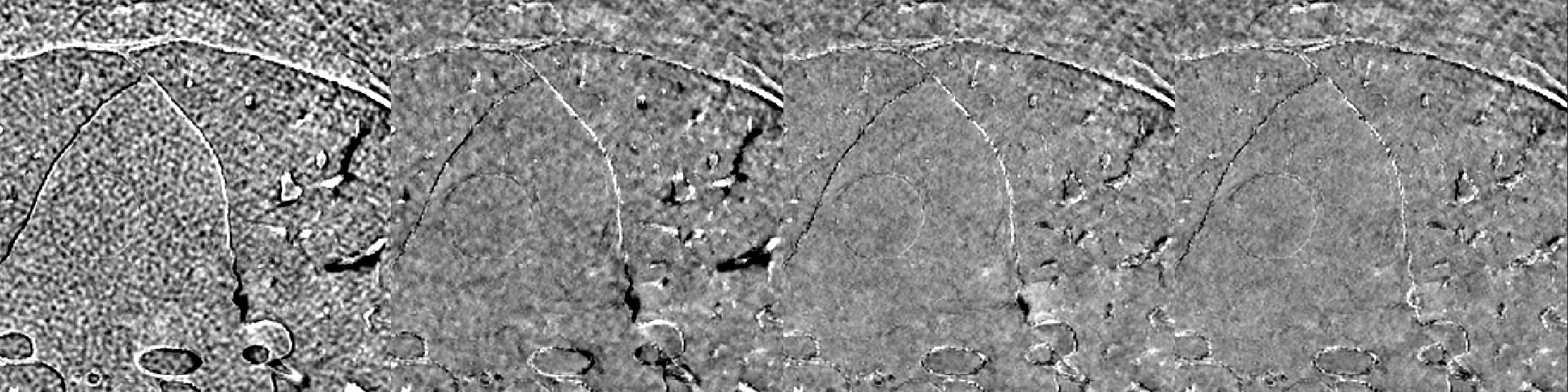}\label{fig:Error_10}}
	\hspace{0.01 cm}
	\subfloat[]{\includegraphics[width=0.45\textwidth]{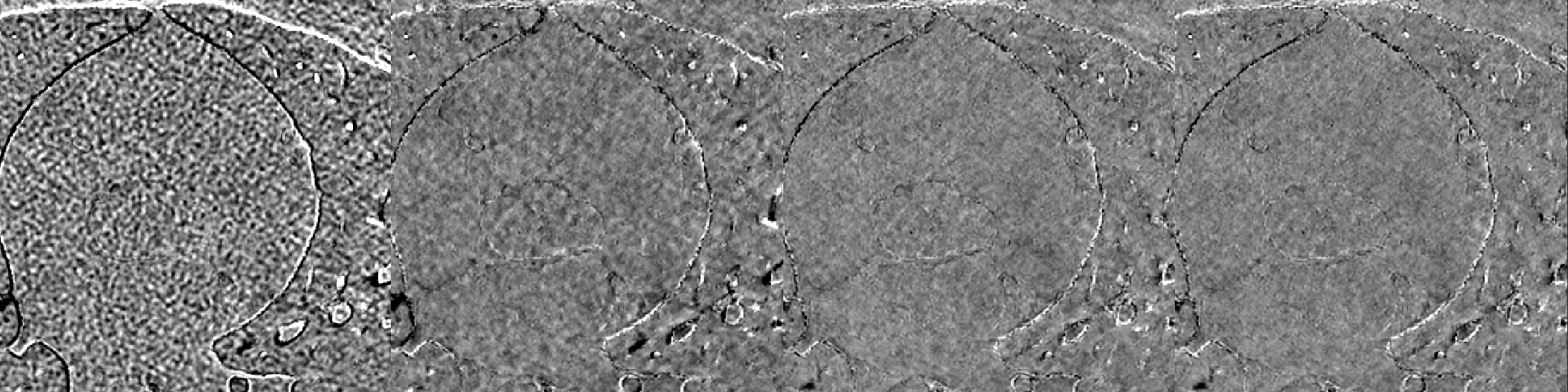}\label{fig:Error_50}}
	\hspace{0.01 cm}
	\subfloat[]{\includegraphics[width=0.45\textwidth]{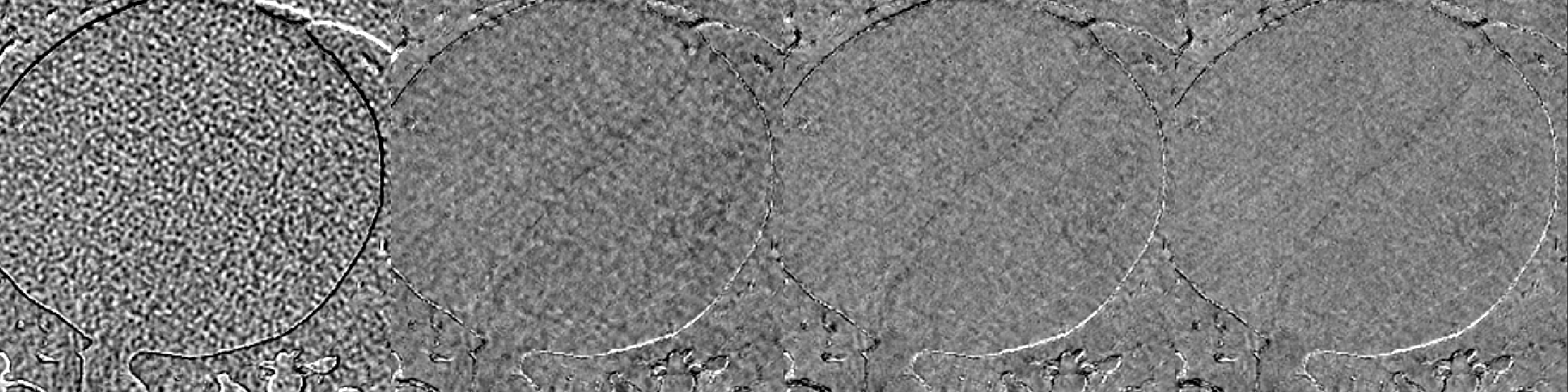}\label{fig:Error_90}}
	\hspace{0.01 cm}
	\subfloat[]{\includegraphics[width=0.45\textwidth]{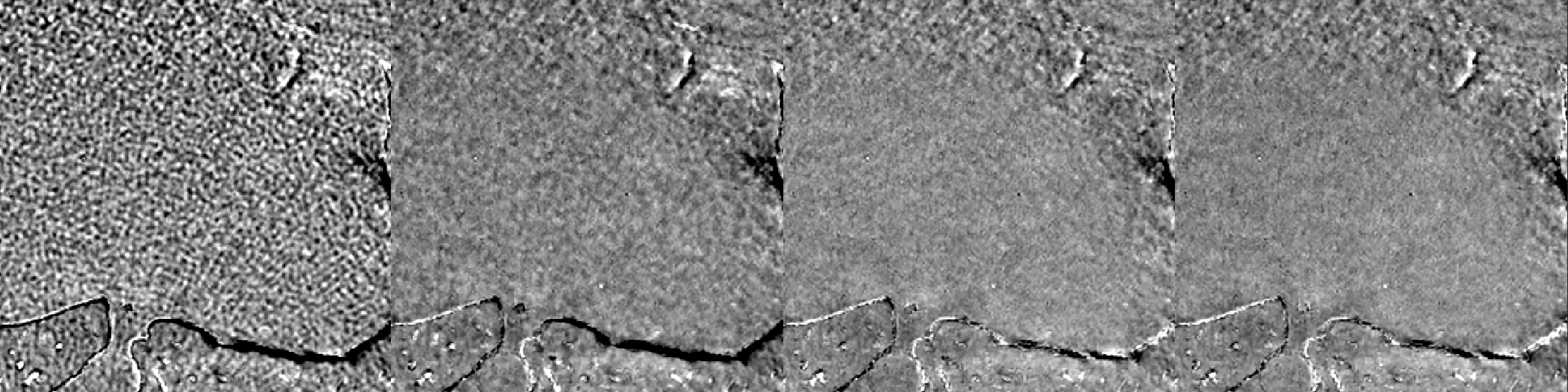}\label{fig:Error_170}}
	\caption{Difference image w.r.t. MBIR. (left to right): $FBP$, $DL \text{-} MBIR_1$, $DL \text{-} MBIR_3$, $DL \text{-} MBIR_5$.
	\label{fig:error}
	\protect\subref{fig:Error_10},  \protect\subref{fig:Error_50},  \protect\subref{fig:Error_90} and \protect\subref{fig:Error_170} represent different slices in the image volume. WW/WL 150/0 HU.}
  \end{center}
\end{figure}

\begin{figure}[h!]
  \begin{center}
	\subfloat[]{\includegraphics[width=0.45\textwidth]{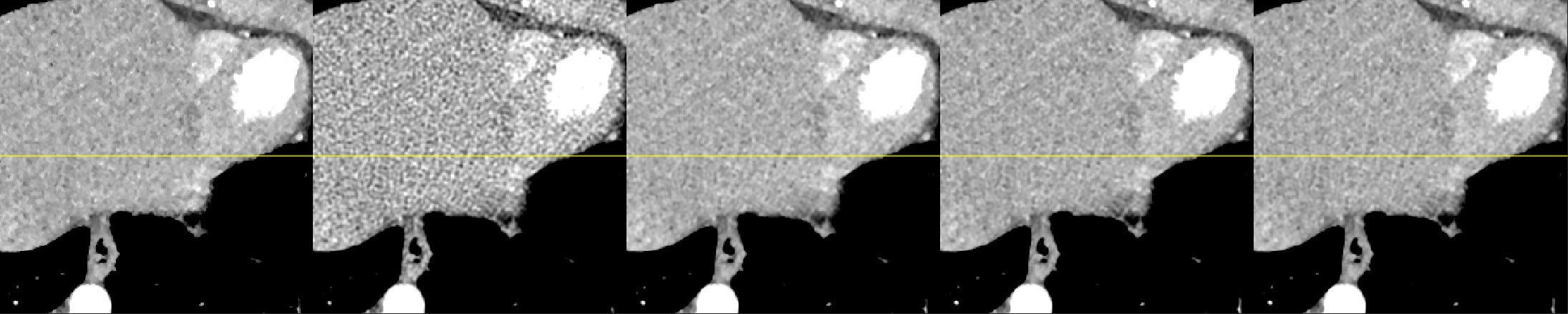}\label{fig:Image_160_yline}}
	\hspace{0.01 cm}
	\subfloat[]{\includegraphics[width=0.45\textwidth]{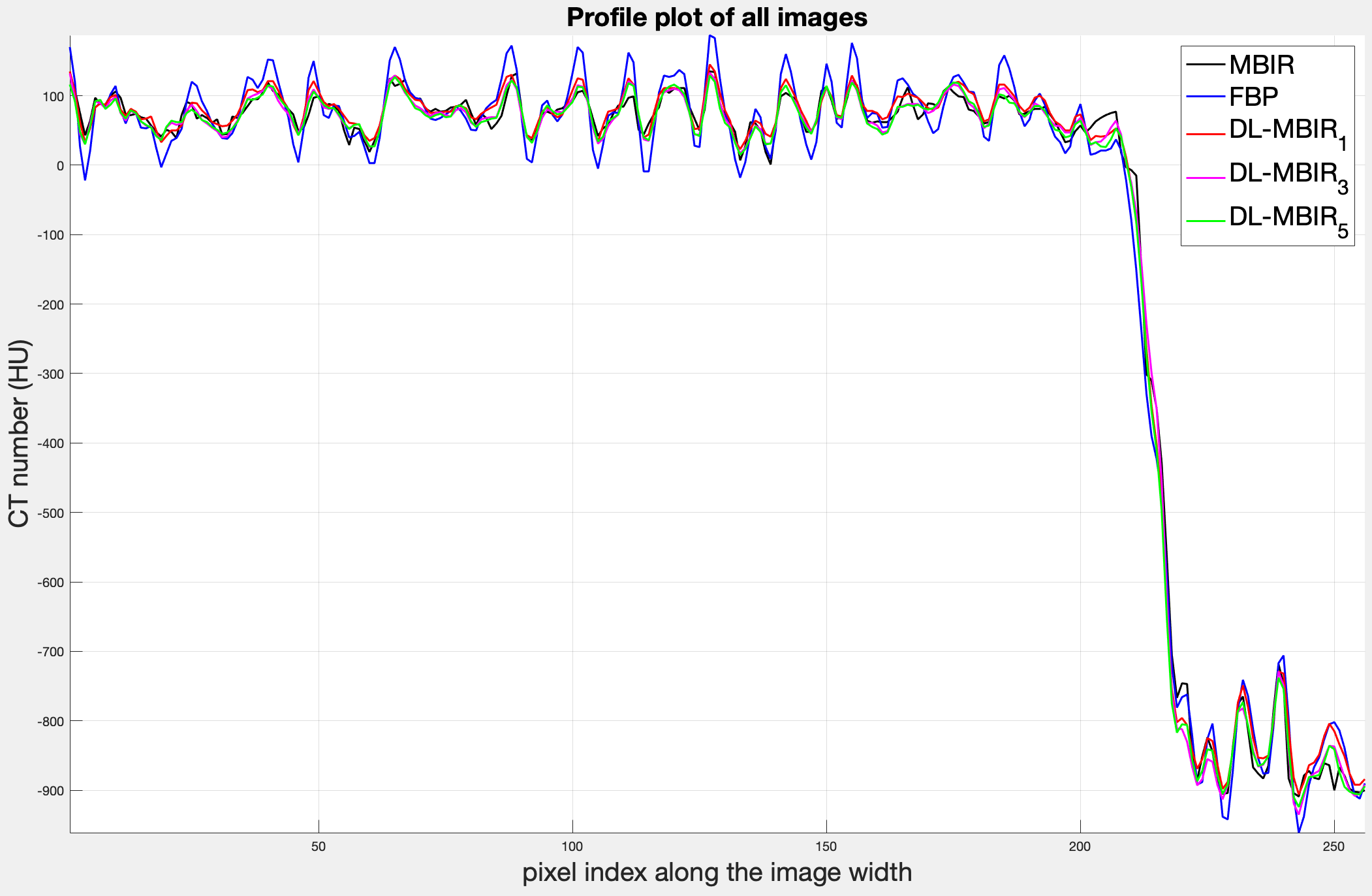}\label{fig:Profile_160}}
	\caption{Image profile.
	\protect\subref{fig:Image_160_yline} One axial slice, from left to right, of: $MBIR$, $FBP$, $DL \text{-} MBIR_1$, $DL \text{-} MBIR_3$, $DL \text{-} MBIR_5$
	\protect\subref{fig:Profile_160} Profile plot of the images along the yellow line.}
	\label{fig:profile plot}
  \end{center}
\end{figure}

A cardiac FBP image was inferenced on the trained DL-MBIR network. Inference time for every network was between 4 and 6 seconds,
and it goes up with the increase in the number of input channels. The MBIR version of the same exam was also available. 
Figure \ref{fig:recon images} shows a comparison, for 4 slices -- \subref{fig:Image_10}, \subref {fig:Image_50}, \subref{fig:Image_90}, and \subref{fig:Image_170}
in the image volume, among MBIR image, FBP image, and the outputs of $DL \text{-} MBIR_Z$, where $Z=1,\ 3,\ 5$.
Figure \ref{fig:error} shows a comparison, for the same slices in the image volume, among difference between images and the MBIR images.
Figure \ref{fig:profile plot} has a profile plot to show the comparison of $DL \text{-} MBIR_Z$ and FBP images w.r.t the MBIR images.

\begin{figure}
  \centering
  \includegraphics[width=0.45\textwidth]{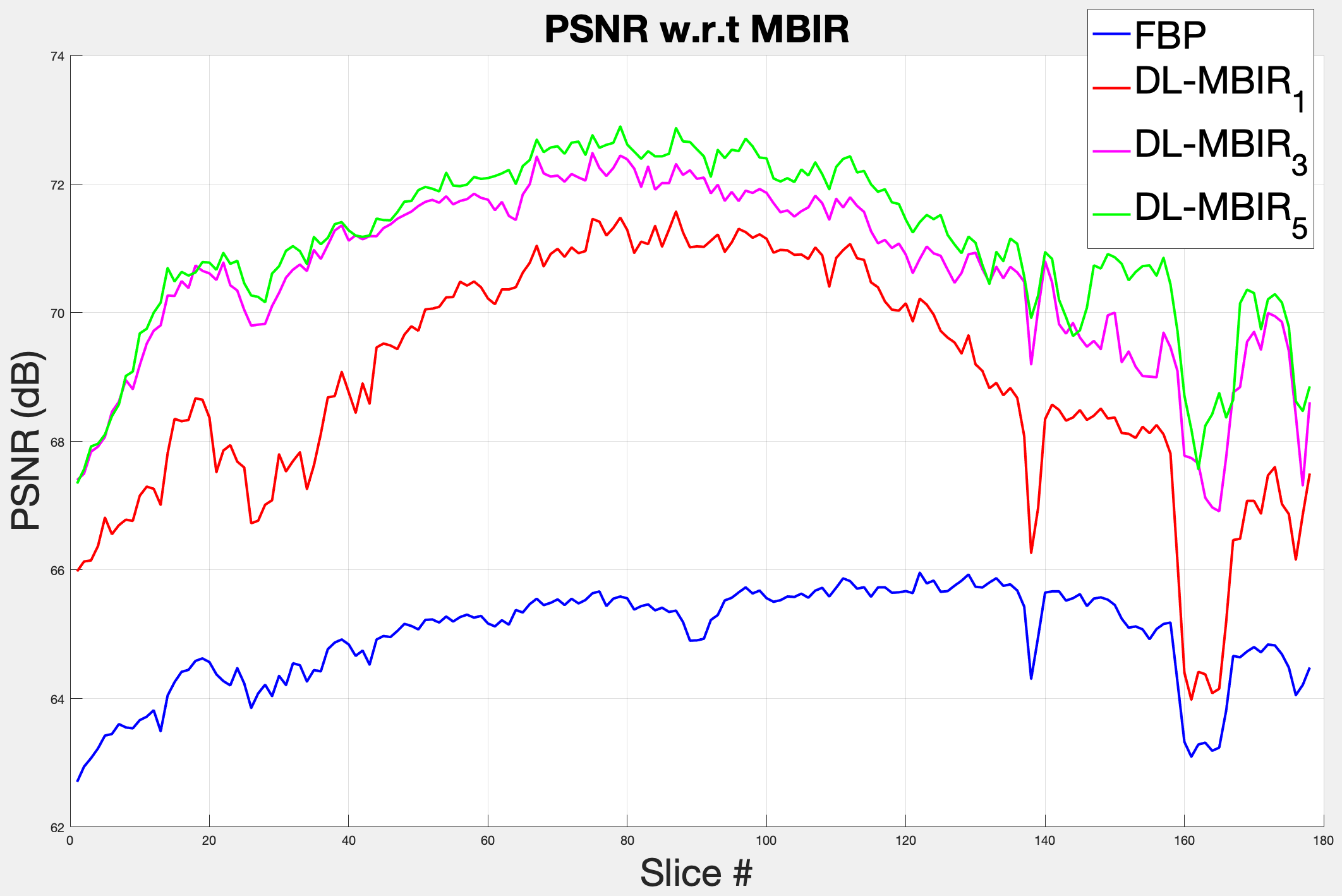}
  \caption{PSNR (with MBIR image as reference). PSNR values are in dB. X axis represents the axial slices in the image volume.}
  \label{fig:PSNR}
\end{figure}

Peak signal to noise ratio (PSNR) is another measure of similarity between images and is closely related to mean squared error.
A higher value would mean that the image is closer to the reference image. 
Figure \ref{fig:PSNR} is a plot of PSNR of all slices within the images with MBIR as the reference image.
Table \ref{tab:Metrics} has some other metrics of comparison among the images, such as averaged (across slices) PSNR, standard deviation
(std) within regions of interest (RoIs), and average CT number within those RoIs.

\begin{figure}
  \centering
  \includegraphics[width=0.45\textwidth]{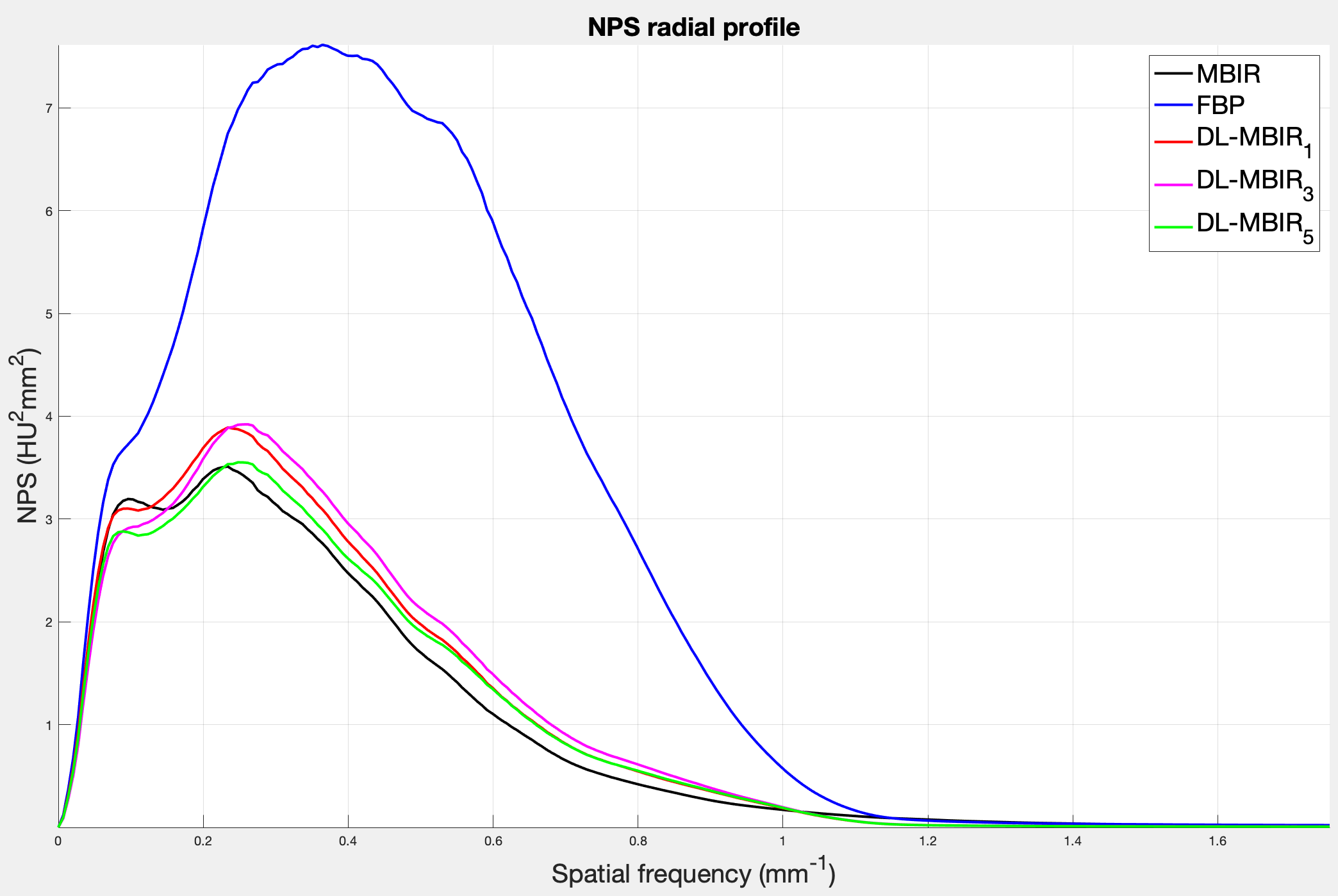}
  \caption{NPS plot demonstrating similar behavior of DL-MBIR with MBIR for noise power.}
  \label{fig:NPS_profile}
\end{figure}

The noise power spectrum (NPS) is a reliable tool for demonstrating similarity in the image texture.
To measure NPS, uniform region patches from one of the cardiac chambers were extracted from each image.
Then NPS was measured for all patches and averaged. Then 1D radial profile was measured from the 2D NPS.
Figure \ref{fig:NPS_profile} shows NPS in the uniform region within one of the cardiac chambers.

\begin{table}
  \centering
  \begin{tabular}{lrrrrr}
  \toprule
  & \footnotesize{$MBIR$}   & \footnotesize{$FBP$}   & \footnotesize{$DL \text{-} MBIR_1$} & \footnotesize{$DL \text{-} MBIR_3$} & \footnotesize{$DL \text{-} MBIR_5$} \\
  \midrule
  \footnotesize{PSNR}    & \footnotesize{-}          & \footnotesize{64.98}          & \footnotesize{68.99}       & \footnotesize{70.63}        & \footnotesize{71.08}        \\
  \footnotesize{std}   & \footnotesize{25.07}         & \footnotesize{46.65}       & \footnotesize{26.39}        & \footnotesize{27.24}         & \footnotesize{25.85}       \\
  \footnotesize{average}   & \footnotesize{74.68}         & \footnotesize{75.20}       & \footnotesize{77.0}        & \footnotesize{71.52}         & \footnotesize{74.43}       \\
  \bottomrule
  \end{tabular}
  \caption{Performance metrics. PSNR (dB) is calculated w.r.t the MBIR images and averaged for all axial slices. Standard deviation (std) and the average value are in HU, with water being 0 and air -1000.}
  \label{tab:Metrics}
\end{table}

\section{Discussion}
Visually, all DL-MBIR images bear close resemblance to the MBIR images in figure \ref{fig:recon images}.
It is confirmed by the difference images in figure \ref{fig:error}.
In the profile plot of Figure \ref{fig:profile plot}, the DL-MBIR profiles closely follow that of MBIR.

All DL-MBIR images have higher PSNR than that of FBP, with $DL \text{-} MBIR_5$ having the best.
Ziabari, et al \cite{ziabari20182} achieved a PSNR gain over FBP of 3.4 dB for $DL \text{-} MBIR_1$, compared to 4.1 dB
with the current implementation.
For $DL \text{-} MBIR_5$, we have improved the result from 4.25 dB to 6.1 dB over FBP.
DL-MBIR images have standard deviations nearly the same as  MBIR, with $DL \text{-} MBIR_5$ outperforming the rest.
The average within the chosen RoI is more or less preserved in all images.

In figure \ref{fig:NPS_profile}, noise power spectrum (NPS) plots of DL-MBIR images are quite  close to that of MBIR, 
indicating that the DL-MBIR image texture is also similar
to that of MBIR, and it appears this attribute is learned well by the network.
Due to its type of adaptive regularization, MBIR may create distinctive texture in the surviving image noise.
The attenuation of noise is a clear gain; however this texture and its effect on low-contrast detectability
may be of concern to some users.

\section{Conclusion}
We trained a U-net, 2.5D DL network that effectively estimates MBIR results from FBP input images.
The computation cost is also signifcantly less than that of MBIR.
All metrics -- NPS, PSNR, standard deviation, profile plots demonstrate that DL-MBIR images have all the
features of MBIR including noise reduction and noise texture.

\printbibliography

\end{document}